\documentclass[ALICE,manyauthors]{cernphprep}
\usepackage[comma,square,numbers,sort&compress]{natbib}
\usepackage{hyperref}
\usepackage{lineno}
\usepackage{xspace}
\usepackage{siunitx}
\usepackage[utf8]{inputenc}
\usepackage[Symbol]{upgreek}
\usepackage[T1]{fontenc}
\DeclareSIUnit\eVc{\eV\per\clight}
\DeclareSIUnit\keVc{\keV\per\clight}
\DeclareSIUnit\MeVc{\MeV\per\clight}
\DeclareSIUnit\MeVcc{\MeV\per\clight\squared}
\DeclareSIUnit\GeVc{\GeV\per\clight}
\DeclareSIUnit\GeVcc{\GeV\per\clight\squared}
\DeclareSIUnit\clight{\text{\ensuremath{c}}}
\DeclareSIUnit[number-unit-product = ]\percent{\char`\%}
\DeclareSIUnit\fm{\femto\metre}
\usepackage{placeins}
\begin{document}
%

\newcommand{\pp}           {pp\xspace}
\newcommand{\ppbar}        {\mbox{$\mathrm {p\overline{p}}$}\xspace}
\newcommand{\XeXe}         {\mbox{Xe--Xe}\xspace}
\newcommand{\PbPb}         {\mbox{Pb--Pb}\xspace}
\newcommand{\pPb}          {\mbox{p--Pb}\xspace}
\newcommand{\AuAu}         {\mbox{Au--Au}\xspace}
\newcommand{\dAu}          {\mbox{d--Au}\xspace}

\newcommand{\snn}          {\ensuremath{\sqrt{s_{\mathrm{NN}}}}\xspace}
\newcommand{\pt}           {\ensuremath{p_{\rm T}}\xspace}
\newcommand{\meanpt}       {$\langle p_{\mathrm{T}}\rangle$\xspace}
\newcommand{\ycms}         {\ensuremath{y_{\rm CMS}}\xspace}
\newcommand{\ylab}         {\ensuremath{y_{\rm lab}}\xspace}
\newcommand{\etarange}[1]  {\mbox{$\left | \eta \right |~<~#1$}}
\newcommand{\yrange}[1]    {\mbox{$\left | y \right |~<~#1$}}
\newcommand{\dndy}         {\ensuremath{\mathrm{d}N_\mathrm{ch}/\mathrm{d}y}\xspace}
\newcommand{\dndeta}       {\ensuremath{\mathrm{d}N_\mathrm{ch}/\mathrm{d}\eta}\xspace}
\newcommand{\avdndeta}     {\ensuremath{\langle\dndeta\rangle}\xspace}
\newcommand{\dNdy}         {\ensuremath{\mathrm{d}N_\mathrm{ch}/\mathrm{d}y}\xspace}
\newcommand{\Npart}        {\ensuremath{N_\mathrm{part}}\xspace}
\newcommand{\Ncoll}        {\ensuremath{N_\mathrm{coll}}\xspace}
\newcommand{\dEdx}         {\ensuremath{\textrm{d}E/\textrm{d}x}\xspace}
\newcommand{\RpPb}         {\ensuremath{R_{\rm pPb}}\xspace}
\newcommand{\mt}           {\ensuremath{m_{\rm T}}\xspace}
\newcommand{\kt}           {\ensuremath{k_{\rm T}}\xspace}
\newcommand{\ks}           {\ensuremath{k^*}\xspace}
\newcommand{\rs}           {\ensuremath{r^*}\xspace}
\newcommand{\ck}           {\ensuremath{C(k^*)}\xspace}
\newcommand{\sr}           {\ensuremath{S(r^*)}\xspace}
\newcommand{\wf}           {\ensuremath{\psi(\vec{\rs},\vec{\ks})}\xspace}
\newcommand{\rc}           {\ensuremath{r_\mathrm{core}}\xspace}
\newcommand{\pP}           {\ensuremath{\mbox{p--p}}\xspace}
\newcommand{\ApAP}         {\ensuremath{\mbox{\pbar--\pbar}}\xspace}
\newcommand{\pL}           {\ensuremath{\mbox{p--\lmb}}\xspace}
\newcommand{\pK}           {\ensuremath{\mbox{p--}\mathrm{K}^-}\xspace}
\newcommand{\ApAL}         {\ensuremath{\mbox{\pbar--\almb}}\xspace}
\newcommand{\LL}           {\ensuremath{\mbox{\lmb--\lmb}}\xspace}
\newcommand{\pSo}          {\ensuremath{\mbox{p--\So}}\xspace}
\newcommand{\pXi}          {\ensuremath{\mbox{p--\X}}\xspace}
\newcommand{\pipi}          {\ensuremath{\uppi\mbox{--}\uppi}\xspace}
\newcommand{\kaka}          {\ensuremath{\mbox{K--K}}\xspace}
\newcommand{\nineH}        {$\sqrt{s}~=~0.9$~Te\kern-.1emV\xspace}
\newcommand{\seven}        {$\sqrt{s}~=~7$~Te\kern-.1emV\xspace}
\newcommand{\onethree}        {$\sqrt{s}~=~13$~Te\kern-.1emV\xspace}
\newcommand{\twoH}         {$\sqrt{s}~=~0.2$~Te\kern-.1emV\xspace}
\newcommand{\twosevensix}  {$\sqrt{s}~=~2.76$~Te\kern-.1emV\xspace}
\newcommand{\five}         {$\sqrt{s}~=~5.02$~Te\kern-.1emV\xspace}
\newcommand{\twosevensixnn}{$\sqrt{s_{\mathrm{NN}}}~=~2.76$~Te\kern-.1emV\xspace}
\newcommand{\fivenn}       {$\sqrt{s_{\mathrm{NN}}}~=~5.02$~Te\kern-.1emV\xspace}
\newcommand{\LT}           {L{\'e}vy-Tsallis\xspace}
\newcommand{\lumi}         {\ensuremath{\mathcal{L}}\xspace}

\newcommand{\ITS}          {\rm{ITS}\xspace}
\newcommand{\TOF}          {\rm{TOF}\xspace}
\newcommand{\ZDC}          {\rm{ZDC}\xspace}
\newcommand{\ZDCs}         {\rm{ZDCs}\xspace}
\newcommand{\ZNA}          {\rm{ZNA}\xspace}
\newcommand{\ZNC}          {\rm{ZNC}\xspace}
\newcommand{\SPD}          {\rm{SPD}\xspace}
\newcommand{\SDD}          {\rm{SDD}\xspace}
\newcommand{\SSD}          {\rm{SSD}\xspace}
\newcommand{\TPC}          {\rm{TPC}\xspace}
\newcommand{\TRD}          {\rm{TRD}\xspace}
\newcommand{\VZERO}        {\rm{V0}\xspace}
\newcommand{\VZEROA}       {\rm{V0A}\xspace}
\newcommand{\VZEROC}       {\rm{V0C}\xspace}
\newcommand{\Vdecay} 	   {\ensuremath{V^{0}}\xspace}

\newcommand{\ee}           {\ensuremath{e^{+}e^{-}}} 
\newcommand{\pip}          {\ensuremath{\uppi^{+}}\xspace}
\newcommand{\pim}          {\ensuremath{\uppi^{-}}\xspace}
\newcommand{\kap}          {\ensuremath{\rm{K}^{+}}\xspace}
\newcommand{\kam}          {\ensuremath{\rm{K}^{-}}\xspace}
\newcommand{\proton}       {\ensuremath{\rm{p}}\xspace}
\newcommand{\pbar}         {\ensuremath{\rm\overline{p}}\xspace}
\newcommand{\kzero}        {\ensuremath{{\rm K}^{0}_{\rm{S}}}\xspace}
\newcommand{\lmb}          {\ensuremath{\Lambda}\xspace}
\newcommand{\lmbs}          {\ensuremath{\Lambda \mathrm{\,baryons}}\xspace}
\newcommand{\almb}         {\ensuremath{\overline{\Lambda}}\xspace}
\newcommand{\So}           {\ensuremath{\Sigma^{0}}\xspace}
\newcommand{\Om}           {\ensuremath{\Omega^-}\xspace}
\newcommand{\Mo}           {\ensuremath{\overline{\Omega}^+}\xspace}
\newcommand{\X}            {\ensuremath{\Xi^-}\xspace}
\newcommand{\Ix}           {\ensuremath{\overline{\Xi}^+}\xspace}
\newcommand{\Xis}          {\ensuremath{\Xi^{\pm}}\xspace}
\newcommand{\Oms}          {\ensuremath{\Omega^{\pm}}\xspace}

\begin{titlepage}
\PHyear{2020}       
\PHnumber{053}      
\PHdate{15 April}  

\title{Search for a common baryon source in high-multiplicity \pp collisions at the LHC}
\ShortTitle{Search for a common particle emitting source}   

\Collaboration{ALICE Collaboration\thanks{See Appendix~\ref{app:collab} for the list of collaboration members}}
\ShortAuthor{ALICE Collaboration} 

\begin{abstract}
We report on the measurement of the size of the particle-emitting source from two-baryon correlations with ALICE in high-multiplicity \pp collisions at \onethree. The source radius is studied with low relative momentum \pP, \ApAP, \pL, and \ApAL pairs as a function of the pair transverse mass \mt considering for the first time in a quantitative way the effect of strong resonance decays. After correcting for this effect, the radii extracted for pairs of different particle species agree. This indicates that protons, antiprotons, \lmb{}s, and \almb{}s originate from the same source. Within the measured \mt range (\num{1.1}--\num{2.2})~\si{\GeVcc} the invariant radius of this common source varies between \num{1.3} and \SI{0.85}{\femto\metre}. These results provide a precise reference for studies of the strong hadron--hadron interactions and for the investigation of collective properties in small colliding systems.

\end{abstract}
\end{titlepage}

\setcounter{page}{2} 


\section{Introduction} 

Correlation techniques have been used in particle physics since the 1960s~\cite{Goldhaber:1960sf}. Significant theoretical progress has been made to relate two-particle correlations at small relative momenta to the study of the space-time properties of the particle-emitting source and the final state interactions between the two particles~\cite{Kopylov:1974th,Lednicky:1981su}. Eventually, these methods were used to study the source size, also referred to as  Hanbury Brown and Twiss (HBT) radius, created in heavy-ion collisions~\cite{Gyulassy:1988yr,Lisa:2005dd,Lisa:2008gf, Henzl:2011dh,Agakishiev:2011zz,Kotte:2004yv,Aggarwal:2007aa,Adams:2004yc,Aamodt:2011mr,Abelev:2014pja,Adam:2015vna}. Collective effects such as hydrodynamic flow introduce position-momentum correlations to the particle emission, and hence modify the source radii in heavy-ion collisions at LHC energies~\cite{Lisa:2005dd}. In these systems, the decrease of the measured source radii with increasing pair transverse momentum $\kt = \mid \vec{p}_{\rm{T},~1} + \vec{p}_{\rm{T},~2}\mid /2$, where $p_{\rm{T}}$ is 
the transverse momentum of each of the particles, and the transverse mass $\mt = \sqrt{\kt^{2} + m^2}$, where $m$ is the average mass of the particle pair, is attributed to the collective expansion of the system created in the collision~\cite{Lisa:2005dd, Bearden2000}. In this context, there are predictions of a common \mt scaling of the radius for different particle pairs, which are based on the assumption of the same flow velocities and freeze-out times for all particle species~\cite{Kisiel:2014upa,Shapoval:2014wya}. There also is experimental evidence that a common \mt scaling of the source radius is present for protons and kaons in heavy-ion collisions~\cite{Adam:2015vja}. On the other hand, for pions the scaling seems to be only approximate~\cite{Adam:2015vja,Adams:2003qa}, which could be explained by the larger effect of the Lorentz boost for lighter particles~\cite{Kisiel:2014upa,Adam:2015vja} but could also be influenced by the effect of feed-down from short-lived resonance decays.
The radii obtained for \PbPb collisions at the LHC can be compared to the freeze-out volume obtained from statistical hadronization models~\cite{Andronic:2017pug} and are also essential ingredients for coalescence models~\cite{Sun:2018mqq,PhysRevC.99.044913,Bellini:2018epz}.

Recent studies of high-multiplicity \pp collisions reveal unexpected similarities to heavy-ion reactions when considering variables normally linked to collective effects, angular correlations, and strangeness production \cite{Khachatryan:2016txc,Khachatryan:2010gv,ALICE:2017jyt, Acharya:2018orn}. The hadronization in \pp collisions is expected to occur on a similar time scale for all particles, and if a common radial velocity for all particles should be present, this would lead to a similar \mt scaling of the source size as measured for heavy-ion collisions. Unfortunately, the information regarding the \mt dependence of the source size measured in \pp collisions is limited to low values of \mt, as the existing data are based on analyses carried out with \pipi and \kaka pairs. These studies point to a variation of the radius as a function of the event multiplicity and of the pair \mt \cite{PhysRevC.97.064912,Aad2015,PhysRevD.87.052016,Abelev:2012ms,CMS:2018vzv}. However, aside a qualitative consideration of a $\beta_\mathrm{T}$ scaling \cite{Humanic:2018mqp}, no quantitative description could be determined so far.

It is known that strongly decaying resonances may lead to significant exponential tails of the source distribution, which can influence in particular the measured \pipi correlations in heavy-ion collisions~\cite{Wiedemann:1996ig,Kisiel:2006is, Sinyukov:2015kga, PhysRevC.81.064906}. This effect is even more pronounced in small collision systems such as \pp and \pPb~\cite{PhysRevD.84.112004,Adam:2015pya}, and can substantially modify the measured source radii, not only for mesons, but for baryons as well. So far a solid modeling of the strong resonance contribution to the source function is still missing.

In this work, we present the first study of the source function with a quantitative evaluation of the effect of strong resonance decays. The search for a common particle-emitting source is conducted employing data measured in high-multiplicity \pp collisions at \onethree. The emission sources of protons and \lmbs are studied using \pP and \pL correlations as a function of the pair \mt. After correcting for the effect of strong resonance decays, the overall source size decreases significantly by up to \SI{20}{\percent} and the values extracted from the different pair combinations are in agreement. The common particle-emitting source described in this work will allow for direct comparisons of the source sizes to the ones resulting from theoretical models and the presence of collective phenomena in small colliding systems to be studied in a complementary way to analyses carried out so far~\cite{PhysRevC.97.064912,Aad2015,PhysRevD.87.052016,Abelev:2012ms,CMS:2018vzv,PhysRevD.84.112004,Adam:2015pya}. These analyses concentrated on \pipi and \kaka correlation studies in \pp collisions, probing the \kt and \mt ranges of up to \si{1}--\SI{1.5}{\GeVcc} and observing a decrease of the source radius at higher \mt, with the measured radii reaching values even below \SI{1}{\femto\metre} in the case of minimum bias events. The higher \mt range is only accessible with baryon femtoscopy.

Additionally, recent ALICE studies revealed that small collision systems, such as \pp, are a suitable environment to study the interaction potential between more exotic pairs, like \pK, \pL, \LL, \pSo, and \pXi~\cite{Acharya:2019bsax,Acharya:2018gyz, FemtoLambdaLambda, FemtoSigma0, Acharya:2019sms}. The data of high-multiplicity triggered \pp collisions at \onethree provides a significantly improved precision compared to the previously analysed minimum bias data. Detailed studies of the interactions will be enabled by a precise knowledge of the size of the common source for particle emission, once corrected for the broadening due to the resonance decays, which depends on the pair type. Moreover, the effective source size is an important input for the modeling of coalescence and has consequences for the prediction of antimatter formation~\cite{PhysRevC.99.044913, Sun:2018mqq, Bellini:2018epz, Blum:2017iwq, Blum:2017qnn}.

\section{Data analysis} 
\label{sec:dataAnalysis}
This paper presents measurements of the \pP, \ApAP, \pL, and \ApAL correlation functions in \linebreak high-multiplicity \pp collisions at \onethree performed with ALICE~\cite{Abelev:2014ffa, 1748-0221-3-08-S08002}. The high-multiplicity trigger selected events based on the measured amplitude in the \VZERO detector system~\cite{Abbas:2013taa}, comprising two arrays of plastic scintillators at $\num{2.8}<\eta<\num{5.1}$ and $\num{-3.7}<\eta<\num{-1.7}$. The threshold was adjusted such that the selected events correspond to the highest \SI{0.17}{\percent} fraction of the multiplicity distribution of all INEL$ > \num{0}$ collisions. In such events, an average of \num{30} charged-particle tracks are found in the range $\mid\eta\mid < \num{0.5}$~\cite{Acharya:2020asf}, which constitutes an increase by a factor of about four with respect to the minimum bias data sample
~\cite{FemtoLambdaLambda}. The \VZERO timing information was evaluated with respect to the LHC clock to distinguish collisions with the beam pipe material or beam--gas interactions. 

The Inner Tracking System (\ITS)~\cite{1748-0221-3-08-S08002} and Time Projection Chamber (\TPC)~\cite{ALME2010316} are the main tracking devices in ALICE. They cover the full azimuthal angle and the pseudorapidity range of $\mid\eta\mid < \num{0.9}$. The solenoid surrounding these detectors creates a homogeneous magnetic field of $B = \SI{0.5}{T}$ directed along the beam axis which defines the $z$ direction. The spatial coordinates of the primary event vertex (PV) are reconstructed once using global tracks reconstructed with the \TPC and \ITS and once using \ITS tracklets~\cite{Abelev:2014ffa}. If both methods yield a vertex, the longitudinal difference between the two, $\Delta z$, is required to be less than \SI{5}{mm}. The $z$ component of the vertex, preferably determined by global tracks, has to lay within $\mid V_{z} \mid < \SI{10}{cm}$ of the nominal interaction point to ensure a uniform detector coverage. Multiple reactions per bunch crossing are identified by the presence of secondary collision vertices~\cite{Abelev:2014ffa}. Approximately \num{e9} events fulfill the above requirements and are available for the analysis. The identification of protons and their respective antiparticles follows the complete set of criteria listed in Refs.~\cite{Acharya:2018gyz,FemtoLambdaLambda}. Primary protons are selected in the transverse- momentum range between \SI{0.5}{\GeVc} and \SI{4.05}{\GeVc} within $\mid\eta\mid<\num{0.8}$. Particle identification (PID) is performed by using the information provided by the \TPC and the Time-Of-Flight (\TOF)~\cite{Akindinov2013} detectors. The energy loss in the \TPC gas is measured for each track, while the timing information of \TOF is required for tracks with $p>$\SI{0.75}{\GeVc}. Particles are identified by a selection on the deviations from the signal hypotheses in units of the respective detector resolution $\sigma_{\TPC}$ and $\sigma_{\TOF}$, according to $n_{\sigma}=\sqrt{n_{\sigma, {\TPC}}^2+n_{\sigma, {\TOF}}^2}<\num{3}$.

The distance of closest approach (DCA) to the PV is restricted to a maximum of \SI{0.1}{\cm} in the transverse plane and \SI{0.2}{\cm} in the $z$ direction, in order to suppress weak decay products or particles created in interactions with the detector material. The composition of the sample is obtained following the methods described in~\cite{Acharya:2018gyz}. For this purpose, events were generated with Pythia 8.2~\cite{Sjostrand:2014zea} (Monash tune~\cite{Skands:2014pea}), processed by GEANT3~\cite{Brun:1987ma}, filtered through the ALICE detector response and subsequently handled by the reconstruction algorithm~\cite{1748-0221-3-08-S08002}. These simulations were used to estimate that the selected protons and antiprotons have a momentum-averaged purity of \SI{99}{\percent}. The fraction of primary and secondary contributions was estimated by a fit of templates of their individual DCA distributions from MC to the \pt-integrated measured distributions. This way the sample was found to consist of \SI{82}{\percent} primary particles. The remainder is due to weak decays of \lmb ($\Sigma^{+}$) baryons contributing with \SI{13}{\percent} (\SI{5}{\percent}).

The \lmb (\almb) candidates are selected following the procedures discussed in \cite{Acharya:2018gyz,FemtoLambdaLambda} by reconstructing the weak decay $\lmb\rightarrow \proton\pim$ ($\almb\rightarrow \pbar\pip$), which has a branching ratio of \SI{63.9}{\percent}~\cite{PhysRevD.98.030001}. The combinatorial background is reduced by requiring the distance of closest approach between the daughter tracks at the secondary vertex to be smaller than \SI{1.5}{cm}. A straight line connecting the secondary vertex with the PV defines the trajectory of the \lmb candidate. Primary \lmbs are selected by requiring a cosine of the pointing angle (CPA) between the momentum vector of the \lmb candidate and its trajectory to be larger than \num{0.99}. The reconstructed daughter particle tracks are required to have an associated hit either in the Silicon Pixel Detector (\SPD) or the Silicon Strip Detector (\SSD) layers of the \ITS or the \TOF detector in order to use their timing information to reduce the remaining contributions from out-of-bunch pile-up. The proton-pion invariant mass distribution is fitted using the sum of a double Gaussian to describe the signal and a second order polynomial for the combinatorial background. In the \pt range between \SI{0.3} to \SI{4.3}{\GeVc}, the \lmb and \almb candidates are reconstructed with a mass resolution between \SI{1.5}{\MeVcc} and \SI{1.8}{\MeVcc}. Choosing a mass window of \SI{4}{\MeVcc} around the nominal mass~\cite{PhysRevD.98.030001} results in a \pt-averaged purity of \SI{96}{\percent}. Similarly to the case of protons, CPA templates of the primary and secondary contributions are generated using MC simulations. These and a production ratio between \lmb and $\Sigma^0$ of 1/3~\cite{200079,ACCIARRI1994223,Adamczewski-Musch:2017gjr,VanBuren:2004hx}, are used to decompose the sample of selected \lmb and \almb candidates. It is found to consist of \SI{59}{\percent} \lmbs directly produced in the collision, while \SI{19}{\percent} originate from electromagnetic decays of a $\Sigma^0$. Additional contributions from weak decays of \X and $\Xi^0$ amount to \SI{11}{\percent} each.

\section{Correlation function}

The observable in femtoscopic measurements is the correlation function \ck, where $\ks=\frac{1}{2}\cdot|\mathbf{p}^*_2-\mathbf{p}^*_1|$ denotes the relative momentum of particle pairs and $\mathbf{p}^*_1 \text{ and }\mathbf{p}^*_2$ are the particle momenta in the pair rest frame (PRF, $\mathbf{p}^*_1 =-\mathbf{p}^*_2$). 
It is computed as $\ck=\mathcal{N} \frac{A(\ks)}{B(\ks)}$, where $A(\ks)$ is the relative momentum distribution of correlated particle pairs, obtained from the same event, and $B(\ks)$ the corresponding distribution of uncorrelated pairs. The latter is obtained by pairing identified particles of one event with particles from a different (``mixed``) event. In order to avoid any bias due to acceptance and reconstruction effects, only those events are mixed, for which the difference between the positions of the vertex in $z$ direction is less than \SI{2}{\cm} and the numbers of global tracks within $\mid\eta\mid < \num{0.8}$ differ by less than four. The normalization factor $\mathcal{N}$ is calculated in the region \ks$\in$ [\num{240},\num{340}]~\si{\MeVc}, where no femtoscopic signal is present and $C(\ks)$ theoretically approaches unity. In the laboratory frame, the single-particle trajectories of \pP and \ApAP pairs at low \ks are almost collinear and hence have a $\Delta\eta$ and $\Delta\varphi^{*}$ $\sim\num{0}$. Here, $\eta$ refers to the pseudorapidity of the track and $\varphi^{*}$ is the azimuthal track coordinate measured at 9 radii in the \TPC, ranging from \SI{85}{\cm} to \SI{245}{\cm}, taking into account track bending because of the magnetic field. Due to detector effects like track splitting and merging~\cite{Adam:2015vja} the reconstruction efficiency for pairs in same and mixed events differs. In order to avoid a bias in the correlation function, a close-pair-rejection (CPR) criterion is applied by removing \pP and \ApAP pairs fulfilling $\sqrt{\Delta\eta^2+\Delta\varphi^{*2}} < \num{0.01}$. For \pL and \ApAL pairs no rejection is considered.

A total number of \num{1.7e6} (\num{1.3e6}) \pP (\ApAP) and \num{0.6e6} (\num{0.5e6}) \pL (\ApAL) pairs are found in the region $\ks<\SI{200}{\MeVc}$. The correlation functions of baryon--baryon pairs agree within statistical uncertainties with their antibaryon\mbox{--}antibaryon pairs~\cite{Adam:2015vja,Adamczyk:2015hza}. Therefore in the following \pP denotes the combination of $\mbox{p--p} \oplus \overline{\mathrm{p}}\mbox{--}\overline{\mathrm{p}}$ and accordingly for \pL. The \pP and \pL correlation functions were obtained separately in \num{7} and \num{6} \mt intervals, respectively, chosen such that the total amount of particle pairs is evenly distributed. 

The theoretical correlation function is related to the two-particle emitting source \sr and wave function \wf ~\cite{Lisa:2005dd}. It can be written as
\begin{equation}
C(\ks)= \int \, \textrm{d}^3 r^* S(r^*) |\wf|^2,
\label{eq:CFsourcewf}
\end{equation}
where $r^*$ is the relative distance between the particle pair defined in the PRF. When fitting this function to the data in this analysis, the free parameters are solely related to \sr. The \wf and the resulting $C(\ks)$ can be determined with the help of the correlation analysis tool using the Schr\"odinger equation (CATS)~\cite{CATS}. The framework was developed in order to model the correlation function in small systems, where the strong interaction can give rise to a particularly pronounced correlation signal. Therefore, \wf is precisely calculated as the numerical solution of the single-channel Schr\"odinger equation, such that additionally to quantum statistics and Coulomb interactions the strong interaction can be included via a local potential $V(\rs)$.

Residual correlations from impurities and feed-down of long-lived resonances decaying weakly or electromagnetically~\cite{Wiedemann:1996ig} are taken into account by calculating the model correlation function $C_{\mathrm{model}}(\ks)$ as
\begin{equation}
C_\mathrm{model}(\ks)=1 + \sum_i \lambda_i ( C_i(\ks) - 1 ) ,    
\label{eq:lambdaCF}
\end{equation}
where the sum runs over all contributions and with the method discussed in Ref.~\cite{Acharya:2018gyz}. In particular the weights $\lambda_{i}$, which are listed separately for \pP and \pL in Table~\ref{tab:lambdaval},  are calculated from purity and feed-down fractions reported in Sec.~\ref{sec:dataAnalysis}.

\begin{table}[!b]
\caption[lambda parameters]{Weight parameters of the individual components of the \pP and \pL correlation function. Misidentifications of particle species X are denoted as $\tilde{\rm{X}}$ and feed-down contributions have the mother particle listed as a sub-index. For the contributions in bold text, the correlation functions are modeled according to the interaction potential, while the others are assumed to be flat. }
\begin{center}
\begin{tabularx}{\textwidth}{l|X || l|X || l|X}
\hline  \hline
\multicolumn{2}{c||}{\pP} & 
\multicolumn{4}{c}{\pL} \\
Pair & $\lambda$ parameter (\si{\percent}) &
Pair & $\lambda$ parameter (\si{\percent}) &
Pair & $\lambda$ parameter (\si{\percent}) \\
\hline 
\boldmath\textbf{pp} & \boldmath\textbf{67.0} &
\boldmath\textbf{p$ \Lambda$} & \boldmath\textbf{46.1} &
p$_{\Sigma^+} \Lambda_{\Xi^0}$ & 0.5 \\

\boldmath\textbf{p$_{\Lambda}$p} & \boldmath\textbf{20.3} &
\boldmath\textbf{p$ \Lambda_{\Xi^{-}}$} & \boldmath\textbf{8.5} &
p$_{\Sigma^+} \Lambda_{\Sigma^0}$ & 1.0\\

p$_{\Lambda}$p$_{\Lambda}$ & 1.5 &
p$ \Lambda_{\Xi^{0}}$ & 8.5 &
$\tilde{\mathrm{p}} \Lambda$ & 0.3  \\
p$_{\Sigma^+}$p & 8.5 &
\boldmath\textbf{p$ \Lambda_{\Sigma^{0}}$} & \boldmath\textbf{15.4} &
$\tilde{\mathrm{p}} \Lambda_{\Xi^-}$ & 0.1 \\

p$_{\Sigma^+}$p$_{\Sigma^+}$ & 0.3 &
p$_{\Lambda} \Lambda$ & 7.0 &
$\tilde{\mathrm{p}} \Lambda_{\Xi^0}$ & 0.1 \\

p$_{\Lambda}$p$_{\Sigma^+}$ & 1.3 &
p$_{\Lambda} \Lambda_{\Xi^-}$ & 1.3 &
$\tilde{\mathrm{p}} \Lambda_{\Sigma^0}$ & 0.1  \\

$\tilde{\mathrm{p}}$p & 0.9 &
p$_{\Lambda} \Lambda_{\Xi^0}$ & 1.3 &
p$ \tilde{\Lambda}$ & 3.3 \\

$\tilde{\mathrm{p}}$p$_{\Lambda}$  & 0.1 &
p$_{\Lambda} \Lambda_{\Sigma^0}$ & 2.3 &
p$_{\Lambda} \tilde{\Lambda}$ & 0.5  \\

$\tilde{\mathrm{p}}$p$_{\Sigma^+}$  & 0.1 &
p$_{\Sigma^+} \Lambda$ &  2.9 &
p$_{\Sigma^+} \tilde{\Lambda}$ & 0.2  \\

$\tilde{\mathrm{p}}\tilde{\mathrm{p}}$ & $0$ &
p$_{\Sigma^+} \Lambda_{\Xi^-}$ & 0.5 &
$\tilde{\mathrm{p}} \tilde{\Lambda}$ & $0$  \\

\hline \hline
\end{tabularx}
\label{tab:lambdaval}
\end{center}
\end{table}

To model the \pP (\pL) correlation function, residual correlations due to the feed-down from \pL (\pSo and \pXi) pairs are explicitly considered, while all other contributions are assumed to be flat. The residual correlations are modeled with CATS assuming the same source radius as the initial particle pair and use theoretical descriptions of their interactions following Ref.~\cite{HAL1,HAL2} for \pXi and Ref.~\cite{Stavinskiy:2007wb,fss2,ESC16} for \pSo. The models describing the \pL interaction will be discussed later in this section. The contributions of these pairs to the \pP and \pL correlation functions have to be scaled by $\lambda_{i}$ and their signal smeared via a decay matrix~\cite{Acharya:2018gyz,PhysRevC.89.054916} which is built according to the kinematics of the decay. Therefore, the residual signal of the initial pair is transformed to the momentum basis of the measured pair. Additionally, each contribution $C_{i}$ is smeared to take into account effects of the finite momentum resolution of the ALICE detector. Except for the genuine correlations, these steps result in a $C_{i}(\ks)\sim\num{1}$ for all combinations, in particular due to the rather small $\lambda$ parameters of most residual contributions as shown in Table~\ref{tab:lambdaval}. Either a constant or a linear baseline $C_{\mathrm{\rm{non-femto}}}(\ks)$ is included in the total fit function $C_{\mathrm{\rm{fit}}}(\ks) = C_{\mathrm{\rm{non-femto}}}(\ks) \cdot C_{\mathrm{\rm{model}}}(\ks)$. The constant factor can, if necessary, introduce a slight correction of the normalization $\mathcal{N}$. The linear baseline function extrapolates any remaining slope of \ck in the normalization region, which may arise due to energy and momentum conservation~\cite{Bock:2011, Acharya:2018gyz}, to the femtoscopic region. The default assumption is a constant, with $C_{\mathrm{\rm{non-femto}}}(\ks) = a$.

The source function \sr is assumed to have a Gaussian profile
\begin{equation}\label{eq:Gauss}
 \sr = \frac{1}{(4\pi r_0 ^2)^{3/2}}\exp{\left(-\dfrac{r^{*2}}{4r_0 ^2}\right)},
\end{equation}
where $r_0$ represents the source radius. The best fit to the \pP correlation function with $C_{\mathrm{\rm{fit}}}(\ks)$ is performed in the region $\ks\in[\num{0},\num{375}]~\si{\MeVc}$ and determines simultaneously all free parameters, namely $r_0$ and the ones related to $C_{\mathrm{\rm{non-femto}}}(\ks)$. The genuine \pP correlation function is calculated by using CATS~\cite{CATS} and the strong Argonne $v_{18}$ potential~\cite{Wiringa:1994wb} in $S$, $P$, and $D$ waves. The systematic uncertainties on $r_{0}$ associated with the fitting procedure are estimated by i) modifying the upper limit of the fit region  to \SI{350}{\MeVc} and \SI{400}{\MeVc}, ii) replacing the normalization $C_{\mathrm{\rm{non-femto}}}(\ks) = a$ ~by a linear function, iii) employing different models describing the residual \pL interaction as discussed later in the text, and iv) modifying the $\lambda$ parameters by varying the composition of secondary contributions by $\pm \SI{20}{\percent}$, while keeping the sum of primary and secondary fractions constant. 

In comparison to \pP, the theoretical models describing the \pL interaction are much less constrained since data from hypernuclei and scattering experiments are scarce~\cite{POLINDER2006244,Haidenbauer:2013oca,Bodmer:1984gc,Acharya:2018gyz,Haidenbauer:2019boi}. The femtoscopic fit is performed in the range $\ks\in[\num{0},\num{224}]~\si{\MeVc}$. The limited amount of experimental data leaves room for different theoretical descriptions of the \pL interaction. In the measurement this is accounted for by performing the fits twice, where the $S$ wave function of the \pL pair is obtained once from chiral effective field theory calculations ($\chi$EFT) at leading order (LO)~\cite{POLINDER2006244} and once from the one at next-to-leading order (NLO)~\cite{Haidenbauer:2019boi}. The systematic uncertainties on $r_{0}$ associated with the fit procedure are estimated by i) changing the upper limit of the fit region to \SI{204}{\MeVc} and \SI{244}{\MeVc}, ii) replacing the normalization constant $C_{\mathrm{\rm{non-femto}}}(\ks) = a$ ~by a linear function, and iii) modifying the $\lambda$ parameters by varying $R_{\Sigma^0/\lmb}$ by $\pm \SI{20}{\percent}$. 

\begin{figure}[!b]
    \centering
    \includegraphics[width=.49\textwidth]{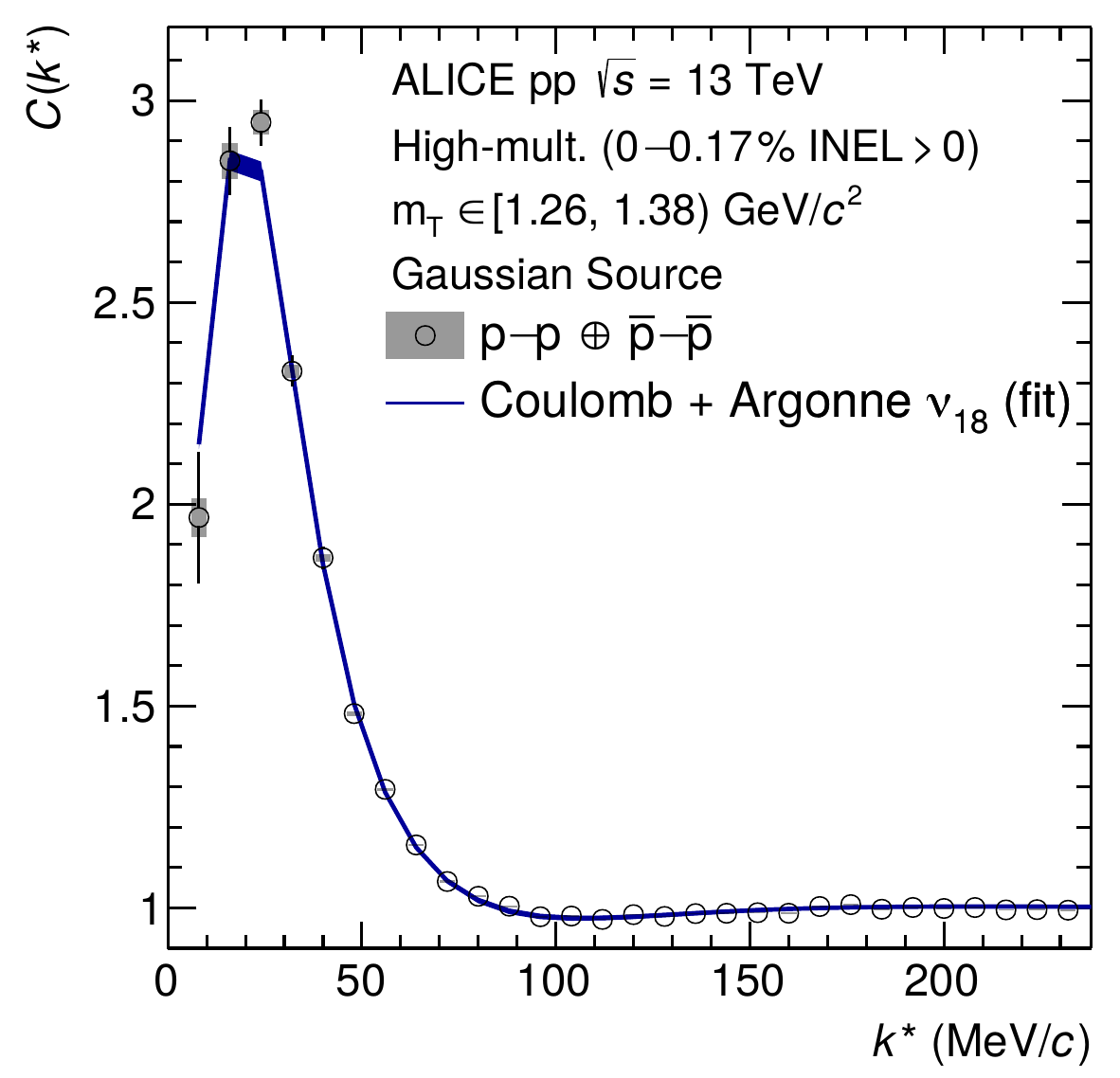}
    \includegraphics[width=.49\textwidth]{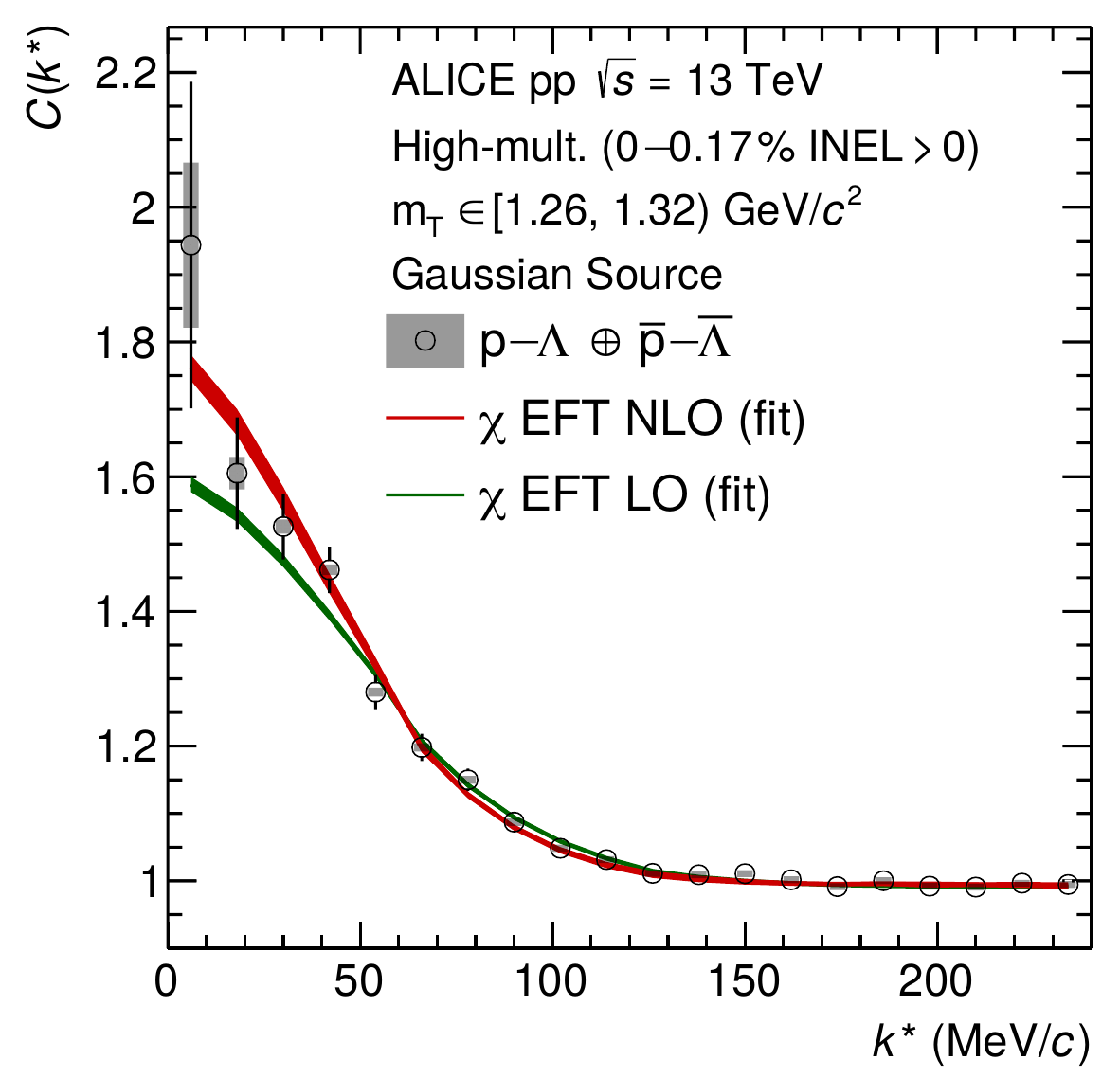}
    \caption{(Color online) The correlation function of \pP (left) and \pL (right) as a function of \ks in one exemplary \mt interval. Statistical (bars) and systematic (boxes) uncertainties are shown separately. The filled bands depict $1\sigma$ uncertainties of the fits with $C_{\mathrm{\rm{fit}}}(\ks)$ and are obtained by using the Argonne $v_{18}$~\cite{Wiringa:1994wb} (blue), $\chi$EFT LO~\cite{POLINDER2006244} (green) and $\chi$EFT NLO~\cite{Haidenbauer:2019boi} (red) potentials. See text for details.}
    \label{fig:mtFit}
\end{figure}

The systematic uncertainties of the experimental \pP and \pL correlation function take into consideration all single-particle selection criteria introduced in the previous section, as well as the CPR criteria on the \pP pairs. All criteria are varied simultaneously up to \SI{20}{\percent} around the nominal values. To limit the  bias of statistical fluctuations, only variations with a maximum change of the pair yield of \SI{20}{\percent} are considered. To obtain the final systematic uncertainty on the source size, the fit procedure is repeated for all variations of the experimental correlation function, using all possible configurations of the fit function. The standard deviation of the resulting distribution for $r_0$ is considered as the final systematic uncertainty.

In Fig.~\ref{fig:mtFit} the \pP and \pL correlation functions of one representative \mt interval are shown. The grey boxes represent the systematic uncertainties of the data and correspond to the $\num{1}\sigma$ interval extracted from the variations of the selection criteria. The resulting relative uncertainty of the \pP (\pL) correlation function reaches a maximum of \SI{2.4}{\percent} (\SI{6.3}{\percent}) in the lowest measured \ks interval. Unlike for meson--meson or baryon--antibaryon pairs, the broad background related to mini-jets is absent for baryon--baryon pairs~\cite{Acharya:2018gyz,Adam:2016iwf}. The width of the fit curves corresponds to the $\num{1}\sigma$ interval extracted from the variations of all the fits. In case of the \pP correlation function, this results in a $\chi^{2}/\textrm{ndf} = \num{1.9}$. The fit of the \pL correlation function using $\chi$EFT calculations at LO yields a $\chi^{2}/\textrm{ndf} = \num{0.91}$ while the fit using $\chi$EFT calculations at NLO yields a $\chi^{2}/\textrm{ndf} = \num{0.67}$. 

\begin{figure}[!b]
    \centering
    \includegraphics[width=1.0\textwidth]{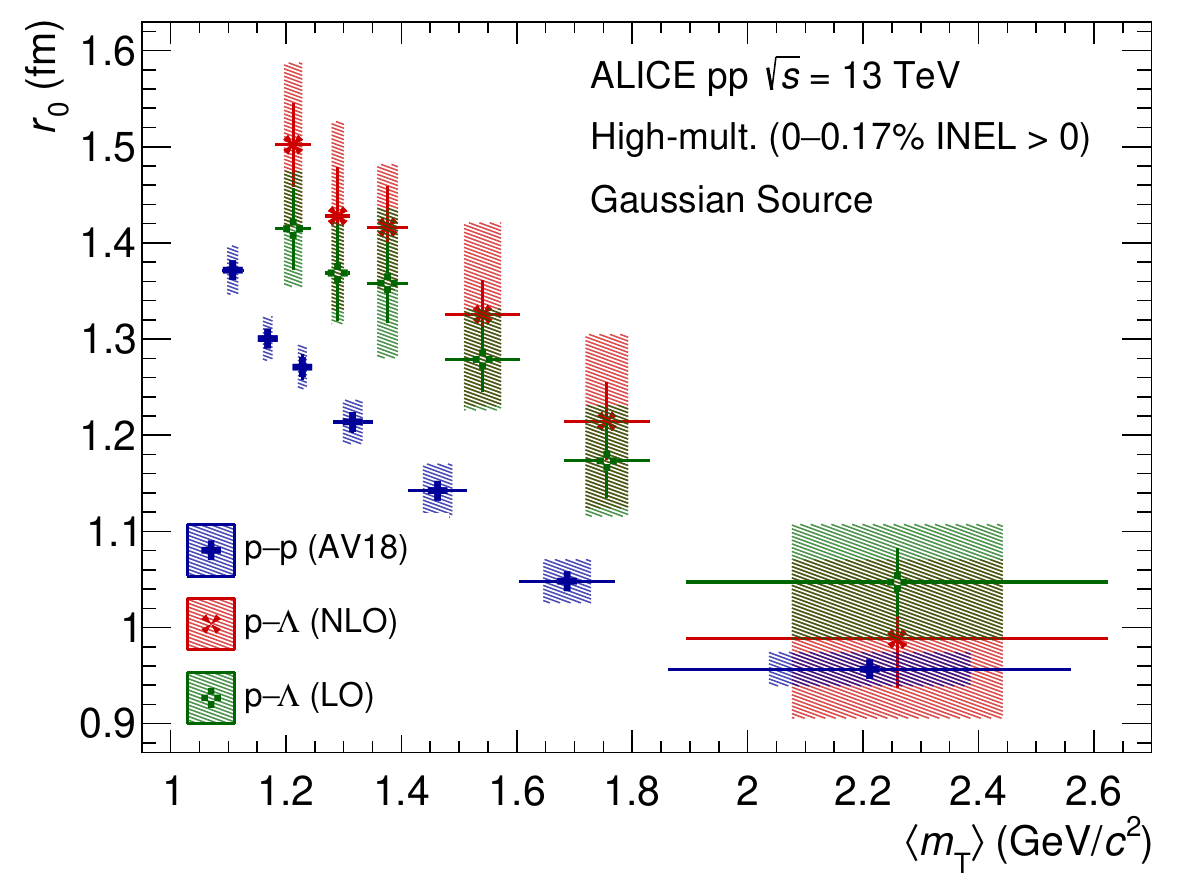}
    \caption{(Color online) Source radius r$_0$ as a function of $\langle \mt \rangle$ for the assumption of a purely Gaussian source. The blue crosses result from fitting the \pP correlation function with the strong Argonne $v_{18}$~\cite{Wiringa:1994wb} potential. The green squared crosses (red diagonal crosses) result from fitting the \pL correlation functions with the strong $\chi$EFT LO~\cite{POLINDER2006244} (NLO~\cite{Haidenbauer:2019boi}) potential. Statistical (lines) and systematic (boxes) uncertainties are shown separately.}
    \label{fig:mtGauss}
\end{figure}

Each correlation function in every \mt interval is fitted and the resulting radii are shown in Fig.~\ref{fig:mtGauss}. The central value corresponds to the mean estimated from the distribution of $r_0$ obtained from the systematic variations. The statistical uncertainties are marked with solid lines, while the boxes correspond to the systematic uncertainties. The relative value of the latter is at most \SI{2.4}{\percent} for the radii extracted from \pP correlations and \SI{8.3}{\percent} and \SI{5.7}{\percent} for those extracted from \pL correlations using the NLO and LO calculations, respectively. The decrease of the source size with increasing \mt is consistent with a hydrodynamic picture, however, the expected common scaling~\cite{Kisiel:2014upa} of the different particle species is not observed for the two considered pair types. The two measurements show a similar trend that is shifted by an offset, indicating that there are differences in the emission of particles.

\section{Modeling the short-lived resonances}

The effect of short-lived resonances ($c\tau \lesssim$ \SI{10}{fm}) feeding into protons and \lmbs could be a possible explanation for the difference between the source sizes determined from \pP and \pL correlations, which was observed in Fig.~\ref{fig:mtGauss}. In the past, Bose-Einstein correlations between identical pions, measured in heavy-ion collisions, were interpreted in terms of a two-component source. It constitutes a core, which is the origin of primary particles, and a halo, which is the origin of pions produced by the decay of resonances~\cite{Csorgo:1994in}. In a detailed investigation of MC simulations of heavy-ion collisions the source sizes were extracted from \pipi pairs for systems both with and without the presence of these contributions, and indeed differences of about \SI{1}{\fm} were found~\cite{PhysRevD.47.3860, Kisiel:2006is}. Similar effects are expected to arise for baryons, since short-lived resonances such as $\Delta$ and N$^*$ decay mainly into a baryon and a pion. The exponential nature of the decay is reflected in the appearance of exponential tails in the source distribution and an effective increase of the source size. Inspired by this picture, a source distribution for baryons is built starting from two components: a Gaussian core and a non-Gaussian halo.

In this work, the resonance yields are taken from the statistical hadronization model (SHM)~\cite{Becattini}. Since this study aims at quantifying the effect of strongly decaying resonances on the source distribution, in the following only primordial particles and secondary decay products of short-lived resonances will be considered. According to the SHM, the amount of primordial protons (\lmbs) are only $P_{\proton}=\SI{35.8}{\percent}$ ($P_{\lmb}=\SI{35.6}{\percent}$)~\cite{BecattiniPrivate}, implying that the effect of the secondaries is substantial. For protons, \num{57} different resonances with lifetimes $\SI{0.5}{fm} < c\tau < \SI{13}{fm}$ are considered. Relative to the total number of protons, \SI{22}{\percent} originate from the decay of a $\Delta^{++}$ resonance, \SI{15}{\percent} from the decay of a  $\Delta^{+}$ resonance, and \SI{7.2}{\percent} from a $\Delta^{0}$ resonance. The remaining secondary protons originate from heavier $\rm{N}^*$, $\Delta$ and \lmb resonances, which contribute individually with less than \SI{2}{\percent}. Similarly, secondary \lmbs stem from \num{32} considered resonances with lifetimes $\SI{0.5}{fm} < c\tau < \SI{8.5}{fm}$. Most prominently $\Sigma^{*+}$, $\Sigma^{*0}$, and $\Sigma^{*-}$ are each the origin of \SI{12}{\percent} of all \lmbs, while decays of heavier $\rm{N}^{*}$, \lmb, and $\Sigma$ resonances individually contribute with less than \SI{1}{\percent}. The weighted average of the lifetimes ($c\tau_\mathrm{res}$) of the resonances feeding into  protons (\lmbs) is \SI{1.65}{fm} (\SI{4.69}{fm}), while the weighted average of the masses is \SI{1.36}{\GeVcc} (\SI{1.46}{\GeVcc}). Although the amount of secondaries is similar for protons and \lmbs, there is a significant difference in the mean lifetime of the corresponding resonances, which is much longer for the \lmb. Qualitatively this will imply a larger effective source size for \pL, as observed in Fig.~\ref{fig:mtGauss}. 

In the following the source function \sr is constructed including the effect of short-lived resonances, assuming that all primordial particles and resonances are emitted from a common Gaussian source of width $r_\mathrm{core}$. Consequently, the particles studied in the final state can either be primordials or decay products of short-lived resonances. For a pair of particles there are four different scenarios regarding their origin, the frequency of each given by $P_1 P_2$, $P_1 \tilde{P}_2$, $\tilde{P}_1 P_2$ and $\tilde{P}_1 \tilde{P}_2$. Here $P_{1,2}$ are the fractions of primordial particles and $\tilde{P}_{1,2}=1-P_{1,2}$ the fractions of particles originating from short-lived resonances. The total source is
\begin{equation}\label{eq:TotalSource}
 \sr = P_1 P_2 \times S_{P_1 P_2}(\rs)+
 P_1 \tilde{P}_2 \times S_{P_1 \tilde{P}_2}(\rs)+
 \tilde{P}_1 P_2 \times S_{\tilde{P}_1 P_2}(\rs)+
 \tilde{P}_1 \tilde{P}_2 \times S_{\tilde{P}_1 \tilde{P}_2}(\rs).
\end{equation}

To evaluate \sr, the required ingredients are the fractions of primordial and secondary particles, and the individual source functions corresponding to the possible combinations for the particle emission. Depending on the average mass and lifetime of the resonances feeding to the particle pair of interest, each of these scenarios will result in slightly different source sizes and shapes. These composite source functions are difficult to compute analytically, however, a simple numerical evaluation, outlined in the following, allows to iteratively build the full source distribution \sr for a given \rc. The primordial emission of particles with a relative distance  $r^*_\mathrm{core}$ is randomly sampled from a Gaussian with width equal to \rc. The resulting particles are then, based on the probabilities $P_{1,2}$ and $\tilde{P}_{1,2}$, assigned to be either primordial particles or resonances. The resonances are propagated and their decays are simulated. For simplicity it is assumed that each decay produces one proton ($\Lambda$) and one pion. It was checked that including three-body decays at this stage would have a negligible effect on the extracted radii. 

\begin{figure}[!ht]
    \centering
    \includegraphics[width=0.75\textwidth]{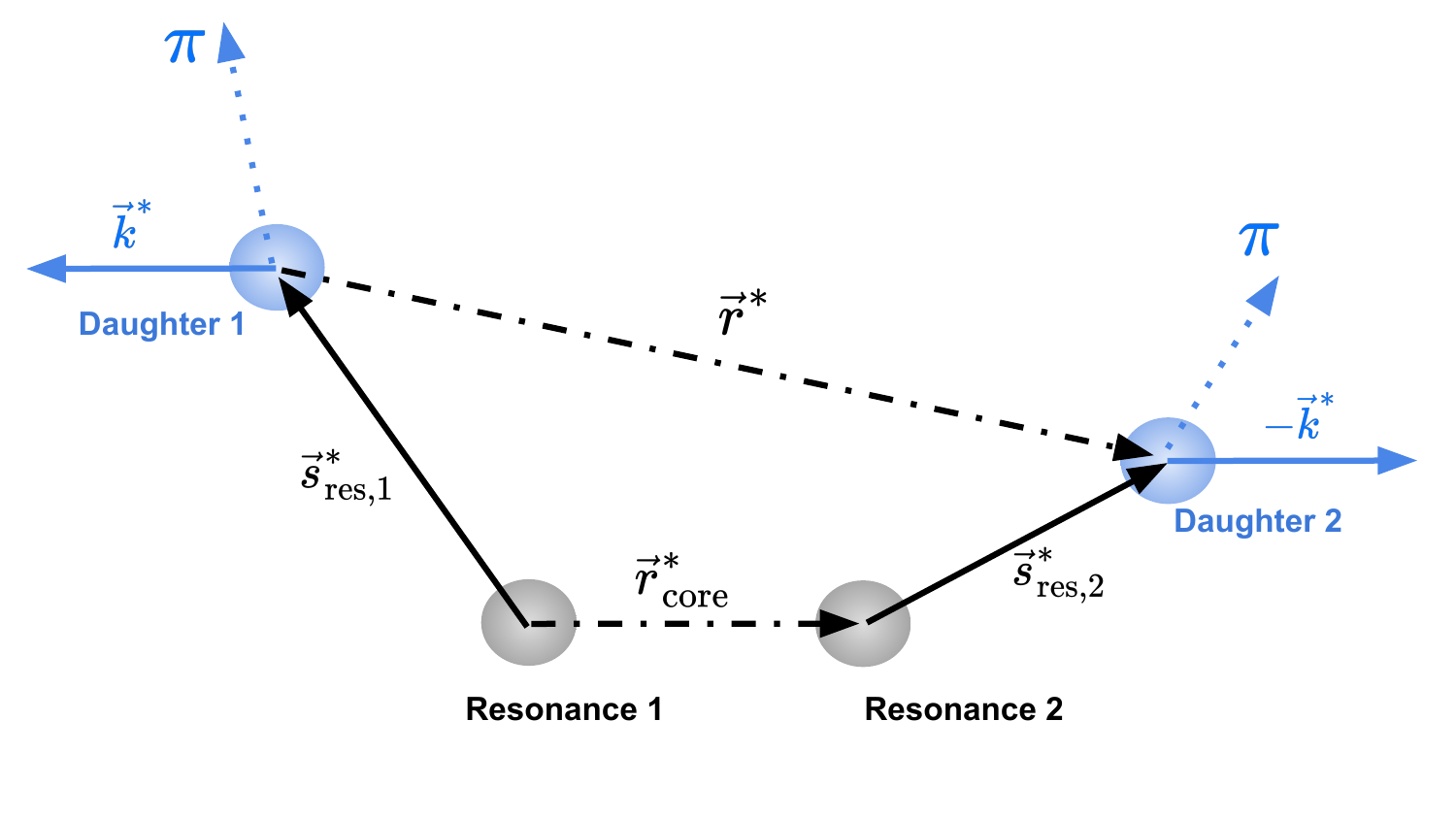}
    \caption{(Color online) A sketch representing the modification of $r^*_\mathrm{core}$ into $r^*$ (dash-dotted lines), due to the presence of resonances (gray disks), decaying into the particles of interest (blue disks). The coordinate system is determined by the rest frame of the two daughters and consistent with Eq.~\ref{eq:CFsourcewf}, where $k^*$ represents their momenta (solid blue lines). The blue dotted lines represent the remaining decay products, which are assumed to be single pions. In case of a primordial particle in the initial state instead of a resonance, the latter is not considered ($\vec{s}^*_\mathrm{res,i}=0$).}
    \label{fig:ResoSketch}
\end{figure}

Figure \ref{fig:ResoSketch} is a schematic representation of the source modification, which in vector form is given as:
\begin{equation}\label{eq:SourceModif}
 \vec{r}^{*}=\vec{r}^{*}_\mathrm{core}-\vec{s}^*_\mathrm{res,1}+\vec{s}^*_\mathrm{res,2},   
\end{equation}
where $\vec{s}^*_\mathrm{res,1(2)}$ is the distance traveled by the first (second) resonance. This is linked to the flight time $t_\mathrm{res}$, which is sampled from an exponential distribution based on the lifetime of the resonance $\tau_\mathrm{res}$:
\begin{equation}\label{eq:Distance}
\vec{s}^*_\mathrm{res}=\vec{\beta}^*_\mathrm{res}\gamma^*_\mathrm{res}t_\mathrm{res}=\frac{\vec{p}^*_\mathrm{res}}{M_\mathrm{res}}t_\mathrm{res},
\end{equation}
where $\vec{p}^*_\mathrm{res}$ is the momentum and $M_\mathrm{res}$ the mass of the corresponding resonance. For the one-dimensional source function $S(r^*)$ the absolute value $r^*=|\vec{r}^*|$ needs to be evaluated. Given the definitions in Eq.~\ref{eq:SourceModif} and Eq.~\ref{eq:Distance}, the required ingredients are $r^*_\mathrm{core}$, the momenta, masses and lifetimes of the resonances, as well as the angles formed by the three vectors $\vec{r}^{*}_\mathrm{core}$, $\vec{s}^*_\mathrm{res,1}$ and $\vec{s}^*_\mathrm{res,2}$. 

The masses and lifetimes of the resonances are fixed to the average values reported above. The remaining unknown parameters, the momenta of the resonances and their relative orientation with respect to $\vec{r}^{*}_\mathrm{core}$, are related to the kinematics of the emission. In this work, the EPOS transport model~\cite{EPOS} is used to quantify these parameters, by generating high-multiplicity \pp events at \onethree and selecting the produced primordial protons, \lmbs and resonances that feed into these particles. Since the yields of the heavier resonances are over-predicted by EPOS, they are weighted such that their average mass $M_\mathrm{res}$ reproduces the expectation from the SHM. The source function \sr is built by selecting a random $r^*_\mathrm{core}$ and a random emission scenario based on the weights $P_{1,2}$, which are known from the SHM. A random EPOS event with the same emission scenario is used to determine $\vec{p}^*_\mathrm{res,1(2)}$ and their relative direction to $\vec{r}^*_\mathrm{core}$. To obtain $r^*$ the resonances are propagated, using Eq. \ref{eq:SourceModif} and \ref{eq:Distance}, and the \ks of their daughters is evaluated. Only events with small \ks are relevant for femtoscopy, thus, if the resulting $k^*>\SI{200}{\MeVc}$, a new EPOS event is picked. The above procedure is repeated until the resulting \sr achieves the desired statistical significance.

\begin{figure}[!ht]
    \centering
    \includegraphics[width=1.\textwidth]{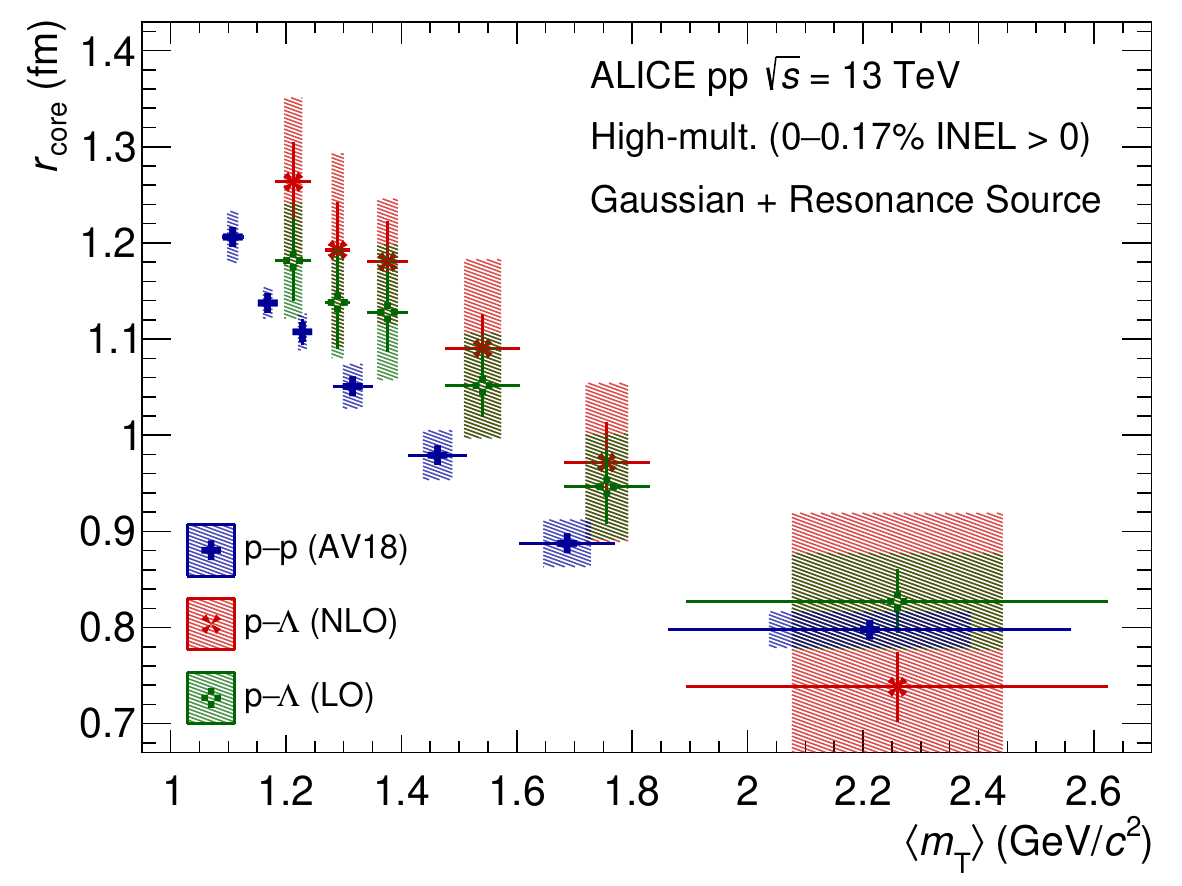}
    \caption{(Color online) Source radius $r_{\textrm{core}}$ as a function of $\langle\mt\rangle$ for the assumption of a Gaussian source with added resonances. The blue crosses result from fitting the \pP correlation function with the strong Argonne $v_{18}$~\cite{Wiringa:1994wb} potential. The green squared crosses (red diagonal crosses) result from fitting the \pL correlation functions with the strong $\chi$EFT LO~\cite{POLINDER2006244} (NLO~\cite{Haidenbauer:2019boi}) potential. Statistical (lines) and systematic (boxes) uncertainties are shown separately. The results shown in this figure were updated with respect to the initial publication~\cite{ALICE:2020ibs} according to App.~\ref{sec:ap}~\cite{Erratum}.}
    \label{fig:mtGaussReso}
\end{figure}

With this method, the modification of the source size due to the decay of resonances is fixed based on the SHM and EPOS, while the only free fit parameter is the size \rc of the primordial (core) source. This procedure is used to refit the \pP and \pL correlation functions. The uncertainties are evaluated in the same way as in the case of the pure Gaussian source. Additional uncertainties due to  short-lived resonances decaying into protons (\lmbs) are accounted for by repeating the fit and altering the mass by \SI{0.2}{\percent} (\SI{0.6}{\percent}) and the lifetimes by \SI{2}{\percent} (\SI{13}{\percent})~\cite{PhysRevD.98.030001}. When comparing the individual fits of the correlation functions in one \mt interval with the ones assuming a pure Gaussian source the resulting $\chi^2$ is found to be similar. This implies that each system can still be described by an effective Gaussian source, albeit loosing the direct physical interpretation of the source size. 
\begin{figure}[!ht]
    \centering
    \includegraphics[width=0.8\textwidth]{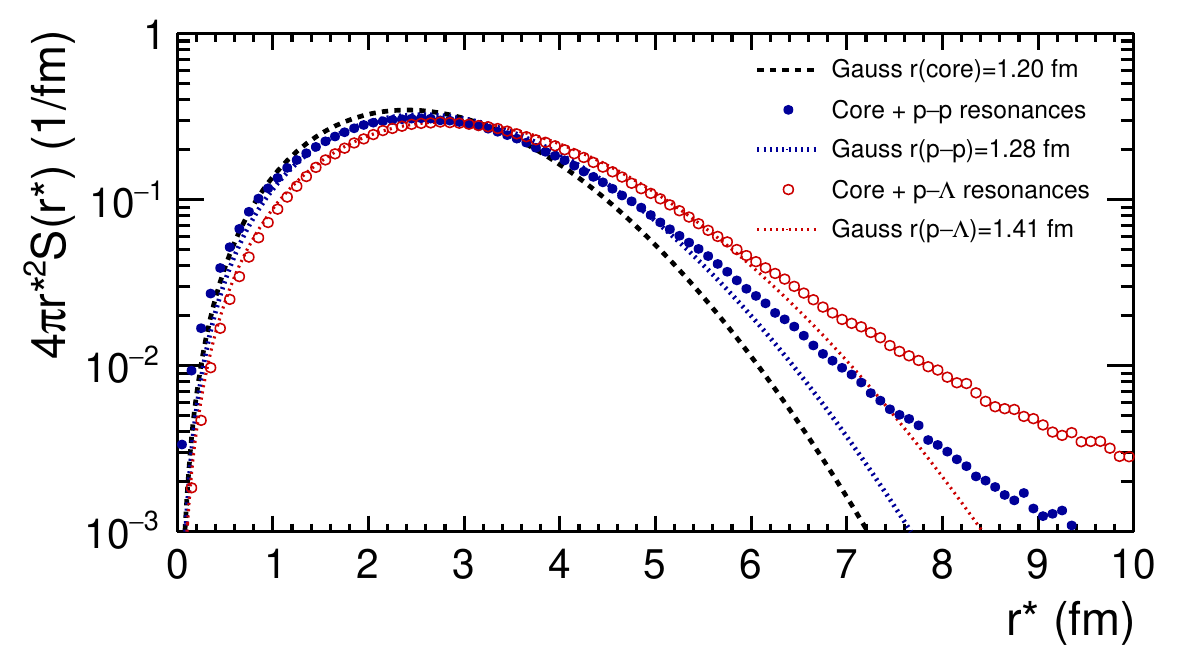}
    \caption{(Color online) The source functions for \pP (blue circles) and \pL (red open circles), generated by folding the exponential expansion due to the decay of the  respective parent resonances with a common Gaussian core with $\rc=\SI{1.2}{fm}$ (dashed black line). Additionally shown are fits with Gaussian distributions (dotted lines) to extract the effective Gaussian source sizes.}
    \label{fig:SourceFunctions}
\end{figure}
This property becomes evident from Fig.~\ref{fig:SourceFunctions}, in which the different source functions, used to describe the \mt bin plotted in Fig.~\ref{fig:mtFit}, are shown. As expected, after the inclusion of the resonances, the same core function results in different effective sources for \pP and \pL. The Gaussian parametrization yields an almost equivalent description of the source function up to about $\rs\sim\SI{6}{fm}$, while for larger values the new parametrization with inclusion of the resonances shows an exponential tail. Since most of the particles are emitted at lower \rs values, the corresponding correlation functions are similar. However, one major difference with the new approach is the resulting source size, as the Gaussian core is more compact than the effective sources. The resulting \mt dependence of \rc measured with \pP and \pL pairs is shown in Fig.~\ref{fig:mtGaussReso}. The relative systematic uncertainty is at most \SI{2.6}{\percent} for the core radii extracted from \pP correlations and \SI{8.4}{\percent} and \SI{6.2}{\percent} for those extracted from \pL correlations using the NLO and LO calculations, respectively. In contrast to a Gaussian source, the new parametrization of the source function provides a common \mt scaling of \rc for both \pP and \pL. This result is compatible with the picture of a common emission source for all baryons and their parent resonances.

\section{Summary} 

The results for \pP and \pL correlations in high-multiplicity \pp collisions at \onethree demonstrate a clear difference in the effective proton and \lmb source sizes if a simple Gaussian source is assumed. A new procedure was developed to quantify for the first time the modification of the source function due to the effect of short-lived resonances. The required input is provided by the statistical hadronization model and the EPOS transport model. The ansatz is that the source function is determined by the convolution of a universal Gaussian core source of size \rc and a non-Gaussian halo. The former represents a universal emission region for all primordial particles and resonances, while the latter is formed by the decay points of the short-lived resonances. This picture is confirmed by the observation of a common \mt scaling of \rc for the \pP and \pL pairs in high-multiplicity \pp collisions, with $\rc\in[\num{0.85},\num{1.3}]$~\si{fm} for $\mt\in[1.1,2.2]~\si{\GeVcc}$. Compared to the values obtained when an effective Gaussian parametrization is used, the overall values are significantly decreased by up to \SI{20}{\percent}.

The measurement of the core size of a common particle-emitting source, corrected for the effect of strong resonances, will allow for direct comparisons with theoretical models. Additionally, detailed studies of the \mt dependence of the core radius will enable complementary investigations of collective phenomena in small collision systems.

On the other hand, the assumption of a common core source, modified by the resonances feeding to the particle pair of interest, allows for a quantitative determination of the effective source for any kind of particle pair. First of all, it enables high-precision studies of the interaction potentials of more exotic baryon--baryon pairs~\cite{Acharya:2018gyz, FemtoLambdaLambda,Acharya:2019sms} that rely on two-particle correlation measurements in momentum space and use the \pP correlation as a reference to fix the emission source. It is also relevant for coalescence approaches addressing the production of (anti) (hyper) nuclear clusters.
A crucial next step is to investigate the applicability of the new method for meson--meson and baryon--meson correlations. If the same \mt scaling is observed as for baryons, this will provide an even more precise quantitative understanding of the common particle-emitting source. In any case, such a study will shed further light on the production mechanism of particles and will be a valuable input for transport models.


\newenvironment{acknowledgement}{\relax}{\relax}
\begin{acknowledgement}
\section*{Acknowledgements}

The ALICE Collaboration would like to thank all its engineers and technicians for their invaluable contributions to the construction of the experiment and the CERN accelerator teams for the outstanding performance of the LHC complex.
The ALICE Collaboration gratefully acknowledges the resources and support provided by all Grid centres and the Worldwide LHC Computing Grid (WLCG) collaboration.
The ALICE Collaboration acknowledges the following funding agencies for their support in building and running the ALICE detector:
A. I. Alikhanyan National Science Laboratory (Yerevan Physics Institute) Foundation (ANSL), State Committee of Science and World Federation of Scientists (WFS), Armenia;
Austrian Academy of Sciences, Austrian Science Fund (FWF): [M 2467-N36] and Nationalstiftung f\"{u}r Forschung, Technologie und Entwicklung, Austria;
Ministry of Communications and High Technologies, National Nuclear Research Center, Azerbaijan;
Conselho Nacional de Desenvolvimento Cient\'{\i}fico e Tecnol\'{o}gico (CNPq), Financiadora de Estudos e Projetos (Finep), Funda\c{c}\~{a}o de Amparo \`{a} Pesquisa do Estado de S\~{a}o Paulo (FAPESP) and Universidade Federal do Rio Grande do Sul (UFRGS), Brazil;
Ministry of Education of China (MOEC) , Ministry of Science \& Technology of China (MSTC) and National Natural Science Foundation of China (NSFC), China;
Ministry of Science and Education and Croatian Science Foundation, Croatia;
Centro de Aplicaciones Tecnol\'{o}gicas y Desarrollo Nuclear (CEADEN), Cubaenerg\'{\i}a, Cuba;
Ministry of Education, Youth and Sports of the Czech Republic, Czech Republic;
The Danish Council for Independent Research | Natural Sciences, the VILLUM FONDEN and Danish National Research Foundation (DNRF), Denmark;
Helsinki Institute of Physics (HIP), Finland;
Commissariat \`{a} l'Energie Atomique (CEA) and Institut National de Physique Nucl\'{e}aire et de Physique des Particules (IN2P3) and Centre National de la Recherche Scientifique (CNRS), France;
Bundesministerium f\"{u}r Bildung und Forschung (BMBF) and GSI Helmholtzzentrum f\"{u}r Schwerionenforschung GmbH, Germany;
General Secretariat for Research and Technology, Ministry of Education, Research and Religions, Greece;
National Research, Development and Innovation Office, Hungary;
Department of Atomic Energy Government of India (DAE), Department of Science and Technology, Government of India (DST), University Grants Commission, Government of India (UGC) and Council of Scientific and Industrial Research (CSIR), India;
Indonesian Institute of Science, Indonesia;
Centro Fermi - Museo Storico della Fisica e Centro Studi e Ricerche Enrico Fermi and Istituto Nazionale di Fisica Nucleare (INFN), Italy;
Institute for Innovative Science and Technology , Nagasaki Institute of Applied Science (IIST), Japanese Ministry of Education, Culture, Sports, Science and Technology (MEXT) and Japan Society for the Promotion of Science (JSPS) KAKENHI, Japan;
Consejo Nacional de Ciencia (CONACYT) y Tecnolog\'{i}a, through Fondo de Cooperaci\'{o}n Internacional en Ciencia y Tecnolog\'{i}a (FONCICYT) and Direcci\'{o}n General de Asuntos del Personal Academico (DGAPA), Mexico;
Nederlandse Organisatie voor Wetenschappelijk Onderzoek (NWO), Netherlands;
The Research Council of Norway, Norway;
Commission on Science and Technology for Sustainable Development in the South (COMSATS), Pakistan;
Pontificia Universidad Cat\'{o}lica del Per\'{u}, Peru;
Ministry of Science and Higher Education, National Science Centre and WUT ID-UB, Poland;
Korea Institute of Science and Technology Information and National Research Foundation of Korea (NRF), Republic of Korea;
Ministry of Education and Scientific Research, Institute of Atomic Physics and Ministry of Research and Innovation and Institute of Atomic Physics, Romania;
Joint Institute for Nuclear Research (JINR), Ministry of Education and Science of the Russian Federation, National Research Centre Kurchatov Institute, Russian Science Foundation and Russian Foundation for Basic Research, Russia;
Ministry of Education, Science, Research and Sport of the Slovak Republic, Slovakia;
National Research Foundation of South Africa, South Africa;
Swedish Research Council (VR) and Knut \& Alice Wallenberg Foundation (KAW), Sweden;
European Organization for Nuclear Research, Switzerland;
Suranaree University of Technology (SUT), National Science and Technology Development Agency (NSDTA) and Office of the Higher Education Commission under NRU project of Thailand, Thailand;
Turkish Atomic Energy Agency (TAEK), Turkey;
National Academy of  Sciences of Ukraine, Ukraine;
Science and Technology Facilities Council (STFC), United Kingdom;
National Science Foundation of the United States of America (NSF) and United States Department of Energy, Office of Nuclear Physics (DOE NP), United States of America.
\end{acknowledgement}

\bibliographystyle{utphys}   
\bibliography{bibliography}

\newpage
\appendix
\section{Appendix}
\label{sec:ap}
The original source code used for evaluating the \rc for \pP pairs contained an error, leading to a slight overestimation of the final result. In this manuscript, this mistake has been corrected, and Fig.~\ref{fig:mtGaussReso} now represents the state-of-the-art evaluation. However, the corrected calculation of \rc introduces a mild tension of 1.9$\sigma$ between the sizes of the common core for \pP and \pL pairs. While statistically insignificant, this deviation has been thoroughly investigated and discussed in an erratum published alongside the journal article associated with this arXiv entry~\cite{Erratum}.  

\begin{figure}[!ht]
    \centering
    \includegraphics[width=1.\textwidth]{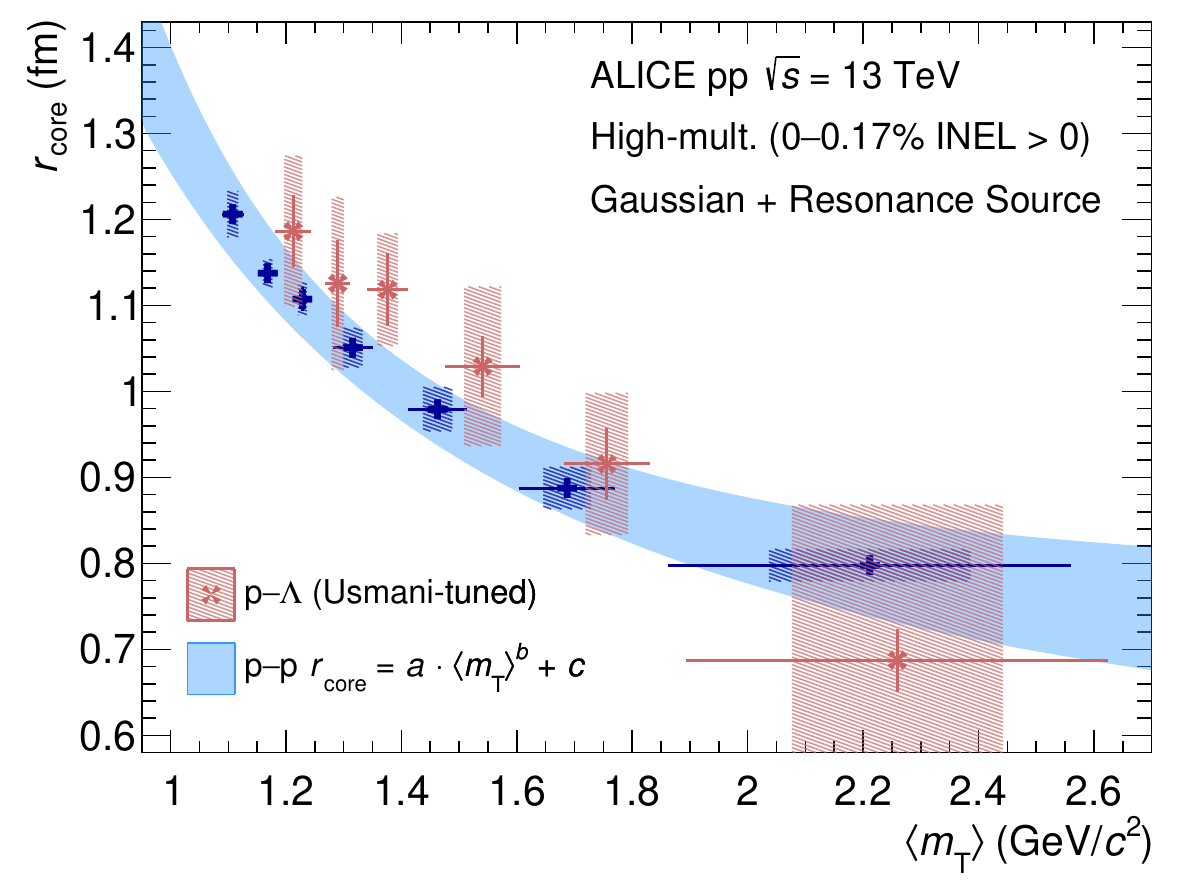}
    \caption{A plot identical to Fig.~\ref{fig:mtGaussReso}, where the red \pL points are obtained by using an Usmani potential tuned to point iv) described in Table 1 in~\cite{Mihaylov:2023ahn}. The blue band is a fit of the \mt dependence to the \pP correlation, details are provided in~\cite{Erratum}.}
    \label{fig:mtGaussResoUsmani}
\end{figure}

Specifically, the \pL correlation functions used in the present work have been reanalyzed using the most recent developments in the modeling of the \pL interaction~\cite{Mihaylov:2023ahn}. In this study the \pL interaction is found to be less attractive compared to the $\chi$EFT tunes done in 2013 and 2019~\cite{Haidenbauer:2013oca,Haidenbauer:2019boi}, leading to a smaller \pL source size (Fig.~\ref{fig:mtGaussResoUsmani}) that is even more consistent with the \pP result ($n_\sigma = 0.90$) and the hypothesis for a common particle-emitting source. 
Further details can be found in~\cite{Erratum} and the references therein.

%
%

\newpage
\section{The ALICE Collaboration}
\label{app:collab}

\begingroup
\small
\begin{flushleft}
S.~Acharya\Irefn{org141}\And 
D.~Adamov\'{a}\Irefn{org95}\And 
A.~Adler\Irefn{org74}\And 
J.~Adolfsson\Irefn{org81}\And 
M.M.~Aggarwal\Irefn{org100}\And 
G.~Aglieri Rinella\Irefn{org34}\And 
M.~Agnello\Irefn{org30}\And 
N.~Agrawal\Irefn{org10}\textsuperscript{,}\Irefn{org54}\And 
Z.~Ahammed\Irefn{org141}\And 
S.~Ahmad\Irefn{org16}
% ...
\And
\renewcommand\labelenumi{\textsuperscript{\theenumi}~}

\section*{Affiliation notes}
\renewcommand\theenumi{\roman{enumi}}
\begin{Authlist}
\item \Adef{org*}Deceased
\item \Adef{orgI}Italian National Agency for New Technologies, Energy and Sustainable Economic Development (ENEA), Bologna, Italy
\item \Adef{orgII}Dipartimento DET del Politecnico di Torino, Turin, Italy
\item \Adef{orgIII}M.V. Lomonosov Moscow State University, D.V. Skobeltsyn Institute of Nuclear, Physics, Moscow, Russia
\item \Adef{orgIV}Department of Applied Physics, Aligarh Muslim University, Aligarh, India
\item \Adef{orgV}Institute of Theoretical Physics, University of Wroclaw, Poland
\end{Authlist}

\section*{Collaboration Institutes}
\renewcommand\theenumi{\arabic{enumi}~}
\begin{Authlist}
\item \Idef{org1}A.I. Alikhanyan National Science Laboratory (Yerevan Physics Institute) Foundation, Yerevan, Armenia
\item \Idef{org2}Bogolyubov Institute for Theoretical Physics, National Academy of Sciences of Ukraine, Kiev, Ukraine
\item \Idef{org3}Bose Institute, Department of Physics  and Centre for Astroparticle Physics and Space Science (CAPSS), Kolkata, India
\item \Idef{org4}Budker Institute for Nuclear Physics, Novosibirsk, Russia
\item \Idef{org5}California Polytechnic State University, San Luis Obispo, California, United States
\item \Idef{org6}Central China Normal University, Wuhan, China
\item \Idef{org7}Centre de Calcul de l'IN2P3, Villeurbanne, Lyon, France
\item \Idef{org8}Centro de Aplicaciones Tecnol\'{o}gicas y Desarrollo Nuclear (CEADEN), Havana, Cuba
\item \Idef{org9}Centro de Investigaci\'{o}n y de Estudios Avanzados (CINVESTAV), Mexico City and M\'{e}rida, Mexico
\item \Idef{org10}Centro Fermi - Museo Storico della Fisica e Centro Studi e Ricerche ``Enrico Fermi', Rome, Italy
\item \Idef{org11}Chicago State University, Chicago, Illinois, United States
\item \Idef{org12}China Institute of Atomic Energy, Beijing, China
\item \Idef{org13}Comenius University Bratislava, Faculty of Mathematics, Physics and Informatics, Bratislava, Slovakia
\item \Idef{org14}COMSATS University Islamabad, Islamabad, Pakistan
\item \Idef{org15}Creighton University, Omaha, Nebraska, United States
\item \Idef{org16}Department of Physics, Aligarh Muslim University, Aligarh, India
\item \Idef{org17}Department of Physics, Pusan National University, Pusan, Republic of Korea
\item \Idef{org18}Department of Physics, Sejong University, Seoul, Republic of Korea
\item \Idef{org19}Department of Physics, University of California, Berkeley, California, United States
\item \Idef{org20}Department of Physics, University of Oslo, Oslo, Norway
\item \Idef{org21}Department of Physics and Technology, University of Bergen, Bergen, Norway
\item \Idef{org22}Dipartimento di Fisica dell'Universit\`{a} 'La Sapienza' and Sezione INFN, Rome, Italy
\item \Idef{org23}Dipartimento di Fisica dell'Universit\`{a} and Sezione INFN, Cagliari, Italy
\item \Idef{org24}Dipartimento di Fisica dell'Universit\`{a} and Sezione INFN, Trieste, Italy
\item \Idef{org25}Dipartimento di Fisica dell'Universit\`{a} and Sezione INFN, Turin, Italy
\item \Idef{org26}Dipartimento di Fisica e Astronomia dell'Universit\`{a} and Sezione INFN, Bologna, Italy
\item \Idef{org27}Dipartimento di Fisica e Astronomia dell'Universit\`{a} and Sezione INFN, Catania, Italy
\item \Idef{org28}Dipartimento di Fisica e Astronomia dell'Universit\`{a} and Sezione INFN, Padova, Italy
\item \Idef{org29}Dipartimento di Fisica `E.R.~Caianiello' dell'Universit\`{a} and Gruppo Collegato INFN, Salerno, Italy
\item \Idef{org30}Dipartimento DISAT del Politecnico and Sezione INFN, Turin, Italy
\item \Idef{org31}Dipartimento di Scienze e Innovazione Tecnologica dell'Universit\`{a} del Piemonte Orientale and INFN Sezione di Torino, Alessandria, Italy
\item \Idef{org32}Dipartimento di Scienze MIFT, Universit\`{a} di Messina, Messina, Italy
\item \Idef{org33}Dipartimento Interateneo di Fisica `M.~Merlin' and Sezione INFN, Bari, Italy
\item \Idef{org34}European Organization for Nuclear Research (CERN), Geneva, Switzerland
\item \Idef{org35}Faculty of Electrical Engineering, Mechanical Engineering and Naval Architecture, University of Split, Split, Croatia
\item \Idef{org36}Faculty of Engineering and Science, Western Norway University of Applied Sciences, Bergen, Norway
\item \Idef{org37}Faculty of Nuclear Sciences and Physical Engineering, Czech Technical University in Prague, Prague, Czech Republic
\item \Idef{org38}Faculty of Science, P.J.~\v{S}af\'{a}rik University, Ko\v{s}ice, Slovakia
\item \Idef{org39}Frankfurt Institute for Advanced Studies, Johann Wolfgang Goethe-Universit\"{a}t Frankfurt, Frankfurt, Germany
\item \Idef{org40}Fudan University, Shanghai, China
\item \Idef{org41}Gangneung-Wonju National University, Gangneung, Republic of Korea
\item \Idef{org42}Gauhati University, Department of Physics, Guwahati, India
\item \Idef{org43}Helmholtz-Institut f\"{u}r Strahlen- und Kernphysik, Rheinische Friedrich-Wilhelms-Universit\"{a}t Bonn, Bonn, Germany
\item \Idef{org44}Helsinki Institute of Physics (HIP), Helsinki, Finland
\item \Idef{org45}High Energy Physics Group,  Universidad Aut\'{o}noma de Puebla, Puebla, Mexico
\item \Idef{org46}Hiroshima University, Hiroshima, Japan
\item \Idef{org47}Hochschule Worms, Zentrum  f\"{u}r Technologietransfer und Telekommunikation (ZTT), Worms, Germany
\item \Idef{org48}Horia Hulubei National Institute of Physics and Nuclear Engineering, Bucharest, Romania
\item \Idef{org49}Indian Institute of Technology Bombay (IIT), Mumbai, India
\item \Idef{org50}Indian Institute of Technology Indore, Indore, India
\item \Idef{org51}Indonesian Institute of Sciences, Jakarta, Indonesia
\item \Idef{org52}INFN, Laboratori Nazionali di Frascati, Frascati, Italy
\item \Idef{org53}INFN, Sezione di Bari, Bari, Italy
\item \Idef{org54}INFN, Sezione di Bologna, Bologna, Italy
\item \Idef{org55}INFN, Sezione di Cagliari, Cagliari, Italy
\item \Idef{org56}INFN, Sezione di Catania, Catania, Italy
\item \Idef{org57}INFN, Sezione di Padova, Padova, Italy
\item \Idef{org58}INFN, Sezione di Roma, Rome, Italy
\item \Idef{org59}INFN, Sezione di Torino, Turin, Italy
\item \Idef{org60}INFN, Sezione di Trieste, Trieste, Italy
\item \Idef{org61}Inha University, Incheon, Republic of Korea
\item \Idef{org62}Institute for Nuclear Research, Academy of Sciences, Moscow, Russia
\item \Idef{org63}Institute for Subatomic Physics, Utrecht University/Nikhef, Utrecht, Netherlands
\item \Idef{org64}Institute of Experimental Physics, Slovak Academy of Sciences, Ko\v{s}ice, Slovakia
\item \Idef{org65}Institute of Physics, Homi Bhabha National Institute, Bhubaneswar, India
\item \Idef{org66}Institute of Physics of the Czech Academy of Sciences, Prague, Czech Republic
\item \Idef{org67}Institute of Space Science (ISS), Bucharest, Romania
\item \Idef{org68}Institut f\"{u}r Kernphysik, Johann Wolfgang Goethe-Universit\"{a}t Frankfurt, Frankfurt, Germany
\item \Idef{org69}Instituto de Ciencias Nucleares, Universidad Nacional Aut\'{o}noma de M\'{e}xico, Mexico City, Mexico
\item \Idef{org70}Instituto de F\'{i}sica, Universidade Federal do Rio Grande do Sul (UFRGS), Porto Alegre, Brazil
\item \Idef{org71}Instituto de F\'{\i}sica, Universidad Nacional Aut\'{o}noma de M\'{e}xico, Mexico City, Mexico
\item \Idef{org72}iThemba LABS, National Research Foundation, Somerset West, South Africa
\item \Idef{org73}Jeonbuk National University, Jeonju, Republic of Korea
\item \Idef{org74}Johann-Wolfgang-Goethe Universit\"{a}t Frankfurt Institut f\"{u}r Informatik, Fachbereich Informatik und Mathematik, Frankfurt, Germany
\item \Idef{org75}Joint Institute for Nuclear Research (JINR), Dubna, Russia
\item \Idef{org76}Korea Institute of Science and Technology Information, Daejeon, Republic of Korea
\item \Idef{org77}KTO Karatay University, Konya, Turkey
\item \Idef{org78}Laboratoire de Physique des 2 Infinis, Irène Joliot-Curie, Orsay, France
\item \Idef{org79}Laboratoire de Physique Subatomique et de Cosmologie, Universit\'{e} Grenoble-Alpes, CNRS-IN2P3, Grenoble, France
\item \Idef{org80}Lawrence Berkeley National Laboratory, Berkeley, California, United States
\item \Idef{org81}Lund University Department of Physics, Division of Particle Physics, Lund, Sweden
\item \Idef{org82}Nagasaki Institute of Applied Science, Nagasaki, Japan
\item \Idef{org83}Nara Women{'}s University (NWU), Nara, Japan
\item \Idef{org84}National and Kapodistrian University of Athens, School of Science, Department of Physics , Athens, Greece
\item \Idef{org85}National Centre for Nuclear Research, Warsaw, Poland
\item \Idef{org86}National Institute of Science Education and Research, Homi Bhabha National Institute, Jatni, India
\item \Idef{org87}National Nuclear Research Center, Baku, Azerbaijan
\item \Idef{org88}National Research Centre Kurchatov Institute, Moscow, Russia
\item \Idef{org89}Niels Bohr Institute, University of Copenhagen, Copenhagen, Denmark
\item \Idef{org90}Nikhef, National institute for subatomic physics, Amsterdam, Netherlands
\item \Idef{org91}NRC Kurchatov Institute IHEP, Protvino, Russia
\item \Idef{org92}NRC «Kurchatov Institute»  - ITEP, Moscow, Russia
\item \Idef{org93}NRNU Moscow Engineering Physics Institute, Moscow, Russia
\item \Idef{org94}Nuclear Physics Group, STFC Daresbury Laboratory, Daresbury, United Kingdom
\item \Idef{org95}Nuclear Physics Institute of the Czech Academy of Sciences, \v{R}e\v{z} u Prahy, Czech Republic
\item \Idef{org96}Oak Ridge National Laboratory, Oak Ridge, Tennessee, United States
\item \Idef{org97}Ohio State University, Columbus, Ohio, United States
\item \Idef{org98}Petersburg Nuclear Physics Institute, Gatchina, Russia
\item \Idef{org99}Physics department, Faculty of science, University of Zagreb, Zagreb, Croatia
\item \Idef{org100}Physics Department, Panjab University, Chandigarh, India
\item \Idef{org101}Physics Department, University of Jammu, Jammu, India
\item \Idef{org102}Physics Department, University of Rajasthan, Jaipur, India
\item \Idef{org103}Physikalisches Institut, Eberhard-Karls-Universit\"{a}t T\"{u}bingen, T\"{u}bingen, Germany
\item \Idef{org104}Physikalisches Institut, Ruprecht-Karls-Universit\"{a}t Heidelberg, Heidelberg, Germany
\item \Idef{org105}Physik Department, Technische Universit\"{a}t M\"{u}nchen, Munich, Germany
\item \Idef{org106}Politecnico di Bari, Bari, Italy
\item \Idef{org107}Research Division and ExtreMe Matter Institute EMMI, GSI Helmholtzzentrum f\"ur Schwerionenforschung GmbH, Darmstadt, Germany
\item \Idef{org108}Rudjer Bo\v{s}kovi\'{c} Institute, Zagreb, Croatia
\item \Idef{org109}Russian Federal Nuclear Center (VNIIEF), Sarov, Russia
\item \Idef{org110}Saha Institute of Nuclear Physics, Homi Bhabha National Institute, Kolkata, India
\item \Idef{org111}School of Physics and Astronomy, University of Birmingham, Birmingham, United Kingdom
\item \Idef{org112}Secci\'{o}n F\'{\i}sica, Departamento de Ciencias, Pontificia Universidad Cat\'{o}lica del Per\'{u}, Lima, Peru
\item \Idef{org113}St. Petersburg State University, St. Petersburg, Russia
\item \Idef{org114}Stefan Meyer Institut f\"{u}r Subatomare Physik (SMI), Vienna, Austria
\item \Idef{org115}SUBATECH, IMT Atlantique, Universit\'{e} de Nantes, CNRS-IN2P3, Nantes, France
\item \Idef{org116}Suranaree University of Technology, Nakhon Ratchasima, Thailand
\item \Idef{org117}Technical University of Ko\v{s}ice, Ko\v{s}ice, Slovakia
\item \Idef{org118}The Henryk Niewodniczanski Institute of Nuclear Physics, Polish Academy of Sciences, Cracow, Poland
\item \Idef{org119}The University of Texas at Austin, Austin, Texas, United States
\item \Idef{org120}Universidad Aut\'{o}noma de Sinaloa, Culiac\'{a}n, Mexico
\item \Idef{org121}Universidade de S\~{a}o Paulo (USP), S\~{a}o Paulo, Brazil
\item \Idef{org122}Universidade Estadual de Campinas (UNICAMP), Campinas, Brazil
\item \Idef{org123}Universidade Federal do ABC, Santo Andre, Brazil
\item \Idef{org124}University of Cape Town, Cape Town, South Africa
\item \Idef{org125}University of Houston, Houston, Texas, United States
\item \Idef{org126}University of Jyv\"{a}skyl\"{a}, Jyv\"{a}skyl\"{a}, Finland
\item \Idef{org127}University of Liverpool, Liverpool, United Kingdom
\item \Idef{org128}University of Science and Technology of China, Hefei, China
\item \Idef{org129}University of South-Eastern Norway, Tonsberg, Norway
\item \Idef{org130}University of Tennessee, Knoxville, Tennessee, United States
\item \Idef{org131}University of the Witwatersrand, Johannesburg, South Africa
\item \Idef{org132}University of Tokyo, Tokyo, Japan
\item \Idef{org133}University of Tsukuba, Tsukuba, Japan
\item \Idef{org134}Universit\'{e} Clermont Auvergne, CNRS/IN2P3, LPC, Clermont-Ferrand, France
\item \Idef{org135}Universit\'{e} de Lyon, Universit\'{e} Lyon 1, CNRS/IN2P3, IPN-Lyon, Villeurbanne, Lyon, France
\item \Idef{org136}Universit\'{e} de Strasbourg, CNRS, IPHC UMR 7178, F-67000 Strasbourg, France, Strasbourg, France
\item \Idef{org137}Universit\'{e} Paris-Saclay Centre d'Etudes de Saclay (CEA), IRFU, D\'{e}partment de Physique Nucl\'{e}aire (DPhN), Saclay, France
\item \Idef{org138}Universit\`{a} degli Studi di Foggia, Foggia, Italy
\item \Idef{org139}Universit\`{a} degli Studi di Pavia, Pavia, Italy
\item \Idef{org140}Universit\`{a} di Brescia, Brescia, Italy
\item \Idef{org141}Variable Energy Cyclotron Centre, Homi Bhabha National Institute, Kolkata, India
\item \Idef{org142}Warsaw University of Technology, Warsaw, Poland
\item \Idef{org143}Wayne State University, Detroit, Michigan, United States
\item \Idef{org144}Westf\"{a}lische Wilhelms-Universit\"{a}t M\"{u}nster, Institut f\"{u}r Kernphysik, M\"{u}nster, Germany
\item \Idef{org145}Wigner Research Centre for Physics, Budapest, Hungary
\item \Idef{org146}Yale University, New Haven, Connecticut, United States
\item \Idef{org147}Yonsei University, Seoul, Republic of Korea
\end{Authlist}
\endgroup
  
\end{document}